\documentclass{moriond}
\usepackage{amsmath}

\def\be{\begin{equation}}
\def\ee{\end{equation}}
\def\bea{\begin{eqnarray}}
\def\eea{\end{eqnarray}}

\newcommand{\Photo}{}

\begin{document}
\vspace*{4cm}
\title{Model independent analysis of heavy vector-like top partners \footnote{Talk given at the 49th Rencontres de Moriond, held in La Thuile (March 2014), and based on \cite{Barducci:2014ila}, carried out in collaboration with D. Barducci, A. Belyaev, G. Cacciapaglia, A. Deandrea, S. De Curtis, J. Marrouche, S. Moretti and L. Panizzi.}}

\author{ M. Buchkremer }

\address{
Centre for Cosmology, Particle Physics and
Phenomenology (CP3), Universit\'e catholique de Louvain, Chemin du
Cyclotron, 2, B-1348, Louvain-la-Neuve, Belgium}

\maketitle
\begin{abstract}
Searches for new fermionic states heavier than the top quark are being
pursued by the CMS \& ATLAS collaborations, pushing the mass bounds towards
the TeV scale. Although a chiral fourth generation of quarks is now excluded by the
LHC results, models going beyond the Standard Model predict the existence of
heavy vector-like top partners as potential smoking gun signatures. Relying on a
model-independent parametrisation, we present the first results of a dedicated
software called \verb|XQCAT| (eXtra Quark Combined Analysis Tool), which
recasts publicly available experimental data from direct and Supersymmetry
inspired searches and computes the exclusion confidence level for New
Physics scenarios with one or multiple top partners. The mass limits set on
a $T$ singlet scenario with general coupling assumptions are briefly
discussed in this framework.
\end{abstract}

\section{Theoretical overview}

Despite the huge success of the Standard Model (SM), compelling arguments indicate that New
Physics (NP) should appear at the TeV\ scale. In particular, the absence of
any symmetries protecting the Brout-Englert-Higgs boson mass term leads to
the expectation that the SM cannot be universally valid.\ As a possible solution to this puzzle, new regimes could be observed at that energy scale, with
novel particles related to the top quark by a symmetry, therefore
carrying similar quantum numbers. These new fermions may then contribute to stabilise the electroweak scale, and cancel the quadratically divergent contributions to the mass of the SM\ scalar boson. 

Although a sequential fourth family of quarks is excluded by the current
searches, models going beyond the SM now point at the possibility for
Vector-Like Quarks (VLQs), \emph{i.e.}, new fermionic resonances having their left-
and right- handed components transforming identically under the electroweak
gauge group. Referred to as \textquotedblleft top partners\textquotedblright
, these hypothetical quarks do not gain their masses from the breaking of
the electroweak symmetry, and occur as a common feature of many NP scenarios such
as extra-dimensional models, strongly interacting dynamics, models
with extended gauge symmetries, Composite Higgs models, and so forth.
Assuming that a single $SU(2)_{L}$ scalar doublet breaks the electroweak symmetry, minimal
scenarios may allow these top partners to interact with the SM quarks through
Yukawa interactions. After that the SM\ scalar boson develops its vacuum
expectation value, the representations summarised in Table \ref{tab:multiplets} are allowed by $SU(2)_{L}\times U(1)_{Y}$ gauge invariance. Vector-like quark singlets and triplets mix dominantly
with the standard left-handed doublets, whereas doublets couple to the
standard right-handed singlets.

\begin{table}[t]
\caption{Vector-like multiplets
have fixed quantum numbers under $SU(2)_{L}\times U(1)_{Y}$, assuming that
they mix with the SM quarks through Yukawa couplings. The electric charge is
the sum of the third component of the isospin $T_{3}$, and of the
hypercharge $Y$.}
\label{tab:multiplets}
\vspace{0.4cm}
\begin{center}
\begin{tabular}{|c||c|c|c|c|c|c|c|}
\hline
$Q_{q}$ & $T_{\frac{2}{3}}$ & $B_{-\frac{1}{3}}$ & $%
\begin{pmatrix}
X_{\frac{5}{3}} \\ 
T_{\frac{2}{3}}%
\end{pmatrix}%
$ & $%
\begin{pmatrix}
T_{\frac{2}{3}} \\ 
B_{-\frac{1}{3}}%
\end{pmatrix}%
$ & $%
\begin{pmatrix}
B_{-\frac{1}{3}} \\ 
Y_{-\frac{4}{3}}%
\end{pmatrix}%
$ & $%
\begin{pmatrix}
X_{\frac{5}{3}} \\ 
T_{\frac{2}{3}} \\ 
B_{-\frac{1}{3}}%
\end{pmatrix}%
$ & $%
\begin{pmatrix}
T_{\frac{2}{3}} \\ 
B_{-\frac{1}{3}} \\ 
Y_{-\frac{4}{3}}%
\end{pmatrix}%
$ \\ \hline\hline
$T_{3}$ & $0$ & $0$ & $1/2$ & $1/2$ & $1/2$ & $1$ & $1$ \\ \hline
$Y$ & $2/3$ & $-1/3$ & $7/6$ & $1/6$ & $-5/6$ & $2/3$ & $-1/3$ \\ \hline
\end{tabular}%
\end{center}
\end{table}

Depending on their charge assignments, these new quarks decay into a standard quark plus a standard gauge boson, such that

\begin{itemize}
\item $X_{5/3} \to W^+ u_i$,

\item $T_{2/3} \to W^+ d_i\,, \; Z u_i\,, \;  H u_i$,

\item $B_{-1/3} \to W^- u_i\,, \; Z d_i\,, \; H d_i$,

\item $Y_{-4/3} \to W^- d_i$,
\end{itemize}

where the index $i=1,2,3$ denotes the three\ standard
generations. The nominal branching fractions for $T$ or $B$ electroweak
singlets are approximately 50\% into $W$ bosons, and 25\% into $Z$ and $%
H $ bosons, in agreement with the Goldstone Boson Equivalence Theorem. The exotic states $X$ and $Y$ decay exclusively through charged currents to up- and down-type quarks, respectively. 

As the guiding thread of this work, we point out that VLQs decaying to top and bottom quarks ($i=3$) have been studied to a great extent at the LHC, while the constraints on top partners decaying to light jets ($i=1,2$) are still mild and require careful treatment. For this reason, we proposed in \cite{Buchkremer:2013bha} a model-independent framework to study the phenomenology
of new top partners at the LHC, with general couplings to all three
generations of SM\ quarks. Factoring out all model-dependent
contributions, we performed a comprehensive analysis of all electroweak production channels for VLQs, relying on a minimal amount of
parameters. Based on this
parametrisation, we provided a compendium of the corresponding cross-section contributions, for benchmark points satisfying the experimental constraints. Our analysis allowed to
highlight the potential relevance of scenarios which have been neglected in
previous experimental searches, as well as of novel interesting channels to
be studied at the LHC.

\section{Reinterpretation of the searches for models with top partners}

Since the start of its physics program, the LHC has delivered a large set of
limits on new heavy coloured objects with spin 1/2. However, reinterpreting
consistently the available bounds for specific NP\ scenarios requires
dedicated strategies to account for all the allowed signals. In the presence of multiple
resonances, events for a given final state may occur through different decay chains
and topologies. Furthermore, the experimental efficiencies may be different
depending on the considered model, affecting the rescaling of the mass
bounds in a non-trivial way.\ Some studies and related codes like CheckMATE, SModelS and Fastlim \cite{Drees:2013wra,Kraml:2013mwa,Papucci:2014rja} have already attempted to tackle this problem, however cannot be thoroughly applied to scenarios with multiple top partners.

In the following we present a dedicated software named \verb|XQCAT|, for \textit{%
eXtra Quark Combined Analysis Tool} \cite{Barducci:2014ila}, that allows the user to determine
the exclusion confidence level for any scenario involving VLQs. Assuming strong pair-production as an input, the corresponding cross-sections are only sensitive to the masses of the new quarks. We simulated $pp\rightarrow Q\bar{Q}+$\{0,1,2\}\mbox{ jets} for each mass value with MadGraph5. The subsequent decays are then computed with BRIDGE, while hadronisation and showering are determined with PYTHIA. Detector simulation is performed through Delphes2. The full signal is
reconstructed by combining, with the appropriate weights, the different
model-independent topologies which generate the underlying final states and
the corresponding kinematic distributions. We subsequently estimate the
number of signal events passing the selection cuts for each signature and
search, and extract the respective efficiencies for each subprocess
contributing to the given final state. Finally, for each implemented search
(or combination thereof), our analysis code evaluates the respective
respective exclusion confidence level for a given input scenario.

To validate and apply our tool, we computed the 95\% CL mass bounds for a $T$
singlet under different hypotheses for its branching ratios into SM bosons
and quarks, considering two selected subsets of searches:  
\begin{enumerate}
\item \textit{Direct searches:} a comprehensive number of final states was
accounted for in \cite{Chatrchyan:2013uxa}.\ In this analysis, the CMS\ collaboration presented an
inclusive search for $Q=2/3$ top partners, at $\sqrt{s}=8$ TeV and 19.5\ fb$%
^{-1}$ of integrated luminosity. Considering pair-produced objects mixing with third-generation quarks, CMS obtained 95\% CL lower limits on $T$
quark masses between 687\ and 782\ GeV. 

\item \textit{SUSY searches: }for the purpose of our analysis, we
implemented the four Supersymmetry-inspired searches \cite{CMS:zxa,Chatrchyan:2012sca,Chatrchyan:2012te,CMS:2012ayl} at $\sqrt{s}=7$ TeV considering the entire
2011 dataset, and characterised by large missing transverse energy and
different numbers of leptons in the final state.\ The updated searches \cite{Chatrchyan:2013lya,Chatrchyan:2013fea} at 8 TeV have been included as well.\ As we have checked that they are
uncorrelated, these searches may be statistically combined.
\end{enumerate}

\begin{figure}[htbp]
\centering
\includegraphics[scale=0.35]{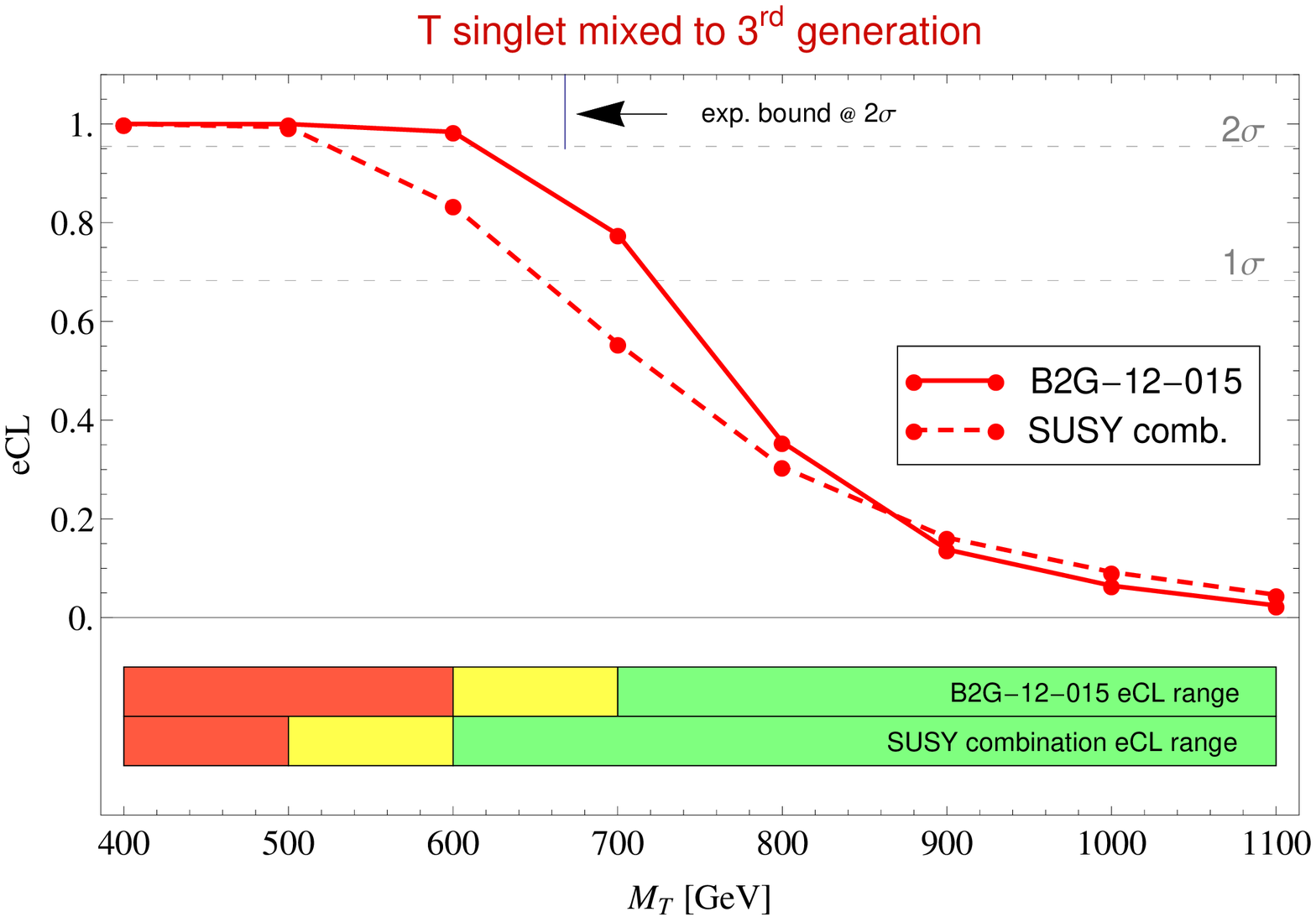}
\includegraphics[scale=0.35]{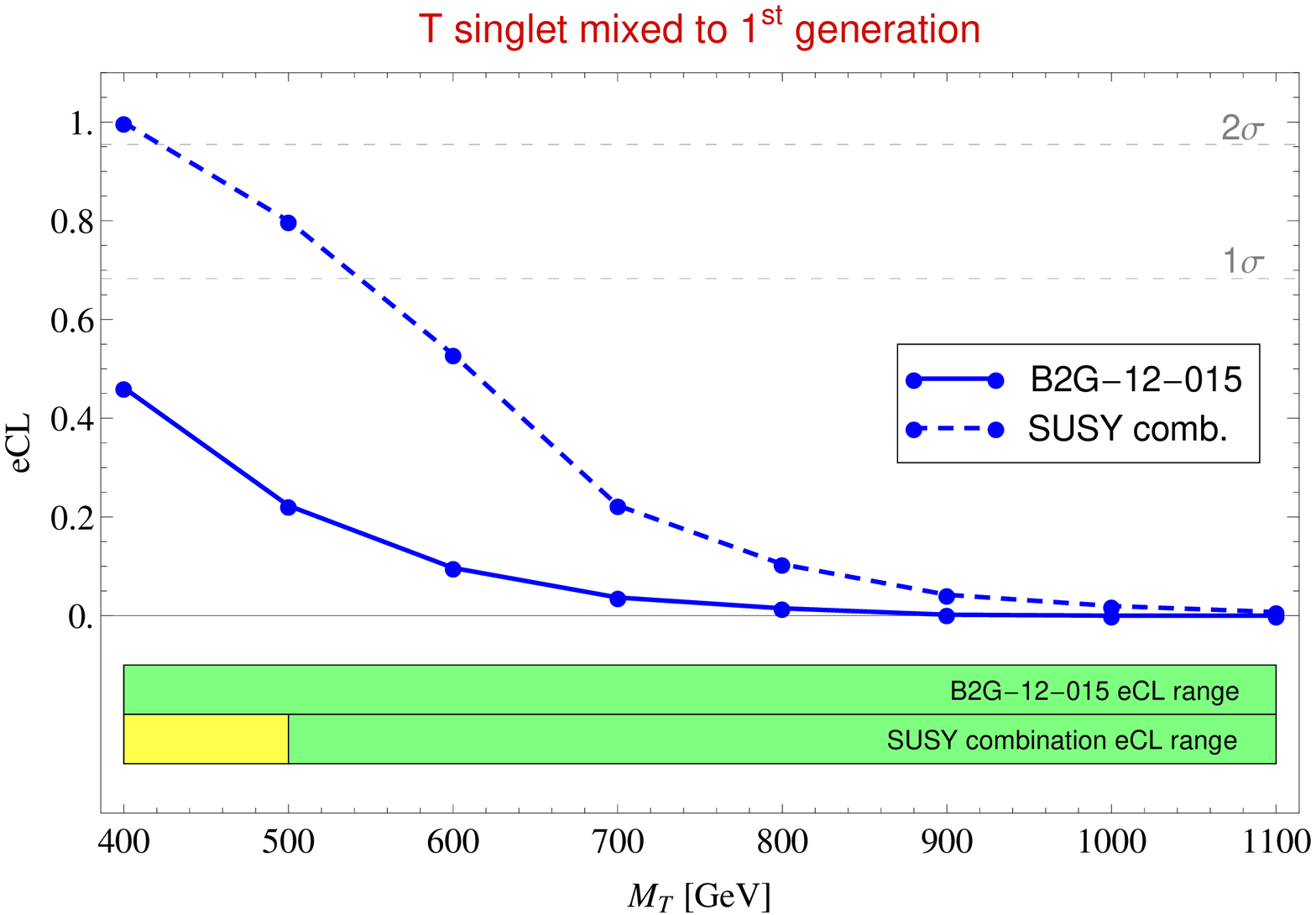}
\caption{Exclusion confidence levels for a $T$ quark mixing only (left) with the SM top quark such that $BR(Wb)=0.5$ and $BR(Zt)=BR(Ht)=0.25$, and (right) to the SM\ up quark such that $BR(Wd)=0.5$ and $BR(Zu)=BR(Hu)=0.25$. The dots correspond to the simulated points, while the lines are linear interpolations of the eCLs. The solid line corresponds to the eCLs obtained using the direct searches, while the dashed line corresponds to the combination of the SUSY searches at $\sqrt{s}=7$ and $8$ TeV. The strips below the plot indicate in red the region where mass values can be excluded at 95\% CL, in yellow the region where excluded mass limits\ can be expected, and in green the region where mass values cannot excluded at the 2 $\sigma$ level.}
\label{fig:T}
\end{figure}

In the left plot of Fig.\ref{fig:T}, we show the exclusion confidence levels\ for a $T$
singlet with $BR(Wb)=50\%$ and $BR(Zt)=BR(Ht)=25\%$. Through linear interpolation of the exclusion confidence levels, we
obtain a 2$\sigma $ mass bound of 614 GeV at 95\%\ CL, whereas the linear interpolation of
the efficiencies excludes masses below 634 GeV, slightly below
the value of 668 GeV quoted in \cite{Chatrchyan:2013uxa} (multilepton channels only). Yet, it is
interesting to notice that the combination of the SUSY\ searches \cite{CMS:zxa,Chatrchyan:2012sca,Chatrchyan:2012te,CMS:2012ayl,Chatrchyan:2013lya,Chatrchyan:2013fea} set constraints in the same ballpark as the direct search \cite{Chatrchyan:2013uxa}.

This selection of searches allows to address an interesting
complementarity between direct analyses for VLQs and others
performed at the LHC. Although no specific search for pair-produced top partners decaying to light jets is currently available, it is shown that
the SUSY searches already set significant bounds on these scenarios.
On the right plot of Fig.\ref{fig:T}, we display the exclusion confidence levels
for a $T$ singlet mixing only with light quarks, such that $BR(Wj)=0.5$
and $BR(Zj)=BR(Hj)=0.25$. The bound provided by linearly interpolating the
eCLs of the SUSY searches combination is 422 GeV (469 GeV if interpolating
the efficiencies), while the direct search \cite{Chatrchyan:2013uxa} does not provide limits above $400$ GeV at the 2$\sigma $ level. Considering that SUSY-inspired searches are not designed to
probe such final states, it is remarkable that their combination is more
sensitive than the analysis \cite{Chatrchyan:2013uxa} for this specific scenario. This points
at the interest of combining multiple topology searches so as to obtain
more accurate bounds on New physics models.

\section{Perspectives}

Details on the ability of \verb|XQCAT| to set bounds on realistic scenarios
with more than one VLQs are given in \cite{Barducci:2014ila}. Future developments
will include further analyses such as recent searches by the ATLAS collaboration,
searches for new bottom partners, and exotic phenomenologies.\ Finally,
adding electroweak single production as a signal input within the context
of our model-independent framework \cite{Buchkremer:2013bha} will allow us to probe the remaining allowed parameter space for VLQs below the TeV scale.

\section*{Acknoweledgements}

I would like to thank the organisers of the \textit{XLIXth Rencontres de
Moriond EW\ 2014} for their invitation and financial support. This work is
supported by the National Fund for Scientific Research (F.R.S.-FNRS) under a
FRIA grant. I also thank Daniele Barducci, Alexander Belyaev, Giacomo
Cacciapaglia, Aldo Deandrea, Stefania De Curtis, Jad Marrouche, Stefano
Moretti and Luca Panizzi for the fruitful collaboration that led to the
results presented in this note.


\section*{References}

\begin{thebibliography}{99}


\bibitem{Barducci:2014ila}
  D.~Barducci, A.~Belyaev, M.~Buchkremer, G.~Cacciapaglia, A.~Deandrea, S.~De Curtis, J.~Marrouche and S.~Moretti {\it et al.},
  arXiv:1405.0737 [hep-ph].
  
  \bibitem{Buchkremer:2013bha}
  M.~Buchkremer, G.~Cacciapaglia, A.~Deandrea and L.~Panizzi,
  Nucl.\ Phys.\ B {\bf 876} (2013) 376
  [arXiv:1305.4172 [hep-ph]].
  
  
\bibitem{Drees:2013wra}
  M.~Drees, H.~Dreiner, D.~Schmeier, J.~Tattersall and J.~S.~Kim,
  arXiv:1312.2591 [hep-ph].
  
\bibitem{Kraml:2013mwa}
  S.~Kraml, S.~Kulkarni, U.~Laa, A.~Lessa, W.~Magerl, D.~Proschofsky-Spindler and W.~Waltenberger,
  arXiv:1312.4175 [hep-ph].
  
\bibitem{Papucci:2014rja}
  M.~Papucci, K.~Sakurai, A.~Weiler and L.~Zeune,
  arXiv:1402.0492 [hep-ph].
  
\bibitem{Chatrchyan:2013uxa}
  S.~Chatrchyan {\it et al.}  [CMS Collaboration],
  Phys.\ Lett.\ B {\bf 729} (2014) 149
  [arXiv:1311.7667 [hep-ex]].
  
\bibitem{CMS:zxa}
  [CMS Collaboration],
  CMS-PAS-SUS-12-028.
  
\bibitem{Chatrchyan:2012sca}
  S.~Chatrchyan {\it et al.}  [CMS Collaboration],
  Phys.\ Rev.\ D {\bf 87} (2013) 5,  052006
  [arXiv:1211.3143 [hep-ex]].
  
\bibitem{Chatrchyan:2012te}
  S.~Chatrchyan {\it et al.}  [CMS Collaboration],
  Phys.\ Lett.\ B {\bf 718} (2013) 815
  [arXiv:1206.3949 [hep-ex]].

\bibitem{CMS:2012ayl}
  CMS Collaboration [CMS Collaboration],
  CMS-PAS-SUS-11-020.

\bibitem{Chatrchyan:2013lya}
  S.~Chatrchyan {\it et al.}  [CMS Collaboration],
  Eur.\ Phys.\ J.\ C {\bf 73} (2013) 2568
  [arXiv:1303.2985 [hep-ex]].
  
\bibitem{Chatrchyan:2013fea}
  S.~Chatrchyan {\it et al.}  [CMS Collaboration],
  JHEP {\bf 1401} (2014) 163
  [arXiv:1311.6736, arXiv:1311.6736 [hep-ex]].
  
  
  

\end{thebibliography}
\end{document}